\pdfoutput=1
\documentclass[linenumbers]{aastex631}
\nolinenumbers
\usepackage{threeparttable}

\newcommand\mearth{M$_\earth$}

\newcommand\msun{M$_\sun$}

\newcommand\rearth{R$_\earth$}
\newcommand\rsun{R$_\sun$}

\newcommand\ben[1]{{\color{black}#1}}

\received{\ben{Dec 09, 2023}}
\revised{\ben{May 13, 2024}}
\accepted{May 29, 2024}
\published{July 26, 2024}
\submitjournal{AJ}

\shorttitle{}
\shortauthors{Liberles et al. 202\ben{4}}

\begin{document}

\title{Variations in the Radius Distribution of Single- and Compact Multiple-transiting Planets}

\correspondingauthor{Benjamin T. Liberles}
\email{bliberles@utexas.edu} 

\author[0009-0003-7437-8743]{Benjamin T. Liberles}
\affiliation{\footnotesize Department of Astronomy, University of Florida, Gainesville, FL 32611, USA}
\affiliation{\footnotesize Department of Astronomy, The University of Texas at Austin, Austin, TX 78712, USA}

\author[0000-0001-7730-2240]{Jason A. Dittmann}
\affiliation{\footnotesize Department of Astronomy, University of Florida, Gainesville, FL 32611, USA}

\author[0000-0001-5730-8485]{Stephen M. Elardo}
\affiliation{\footnotesize Department of Geological Sciences, University of Florida, Gainesville, FL 32611, USA}

\author[0000-0002-3247-5081]{Sarah Ballard}
\affiliation{\footnotesize Department of Astronomy, University of Florida, Gainesville, FL 32611, USA}

\begin{abstract}
\nolinenumbers


Previous work has established the enhanced occurrence of compact systems of multiple small exoplanets around metal-poor stars. Understanding the origin of this effect in the planet formation process is a topic of ongoing research. Here we consider the radii of planets residing in systems of multiple transiting planets, compared to those residing in single-transiting systems, with a particular focus on late-type host stars. We investigate whether the two radius distributions are consistent with being drawn from the same underlying planetary population. 
We construct a planetary sample of 290 planets around late K and M dwarfs containing 149 planets from single-transiting planetary systems and 141 planets from multi-transiting compact multiple planetary systems (54 compact multiples). We performed a two-sample Kolmogorov-Smirnov test, Mann-Whitney U test, and Anderson-Darling k-sampling test on the radius distributions of our two samples. We find statistical evidence (p $<$ 0.0026) that planets in compact multiple systems are larger, on average, than their single-transiting counterparts \ben{for planets with $R_p <$ 6 \rearth}. We determine that the offset cannot be explained by detection bias. We investigate whether this effect could be explained via more efficient outgassing of a secondary atmosphere in compact multiple systems due to the stress and strain forces of interplanetary tides on planetary interiors. We find that this effect is insufficient to explain our observations without significant enrichment in H$_2$O compared to Earth-like bulk composition.

\end{abstract}

\section{Introduction}







Low mass stars (K and M dwarfs) are the most common type of stars and most common planet hosting stars in the Galaxy \citep{Reyle21, Swift13, Henry04}. The \textit{Kepler} mission observed a single $\sim$100 $\text{deg}^2$ region of the sky for 3.5 years to detect transiting planets around other host stars \citep{Borucki10}. \textit{Kepler} discovered thousands of planets which provided the first large scale statistical population of exoplanets. Early statistical \ben{studies} from the \textit{Kepler} mission were focused on measuring the planet occurrence rate as a function of planetary radius and planetary period \citep{Howard12, Dressing13, Hsu19, JiaYi20}. M dwarfs have 3.5 times more small planets ($1.0-2.8$ \rearth) than main-sequence FGK stars \citep{Mulders15}. Aside from raw occurrence rates, the \textit{Kepler} sample also displays a variety of orbital architectures in its planetary sample. \ben{The term ``\textit{Kepler} dichotomy” \citep{Lissauer11} has previously been employed to describe a seeming bimodality in the dynamical temperature of planetary systems \citep{Tremaine15}. For low-mass stars, \cite{Ballard16} uncovered some support for a mixture model of planet occurrence.} While the dynamically cool, compact, and coplanar systems of planets reside around 21$^{+7}_{-5}$\% of early and mid-M dwarfs \citep{Muirhead15}, the transit multiplicity distribution cannot be explained solely with this template. Rather, a second population of dynamically hotter systems is required, to furnish sufficient singly transiting systems to match the observed yield \citep{Ballard16}. Subsequent studies found that a smooth distribution in dynamical temperature, rather than a bimodal distribution, can also reproduce the observed transit multiplicity \citep{Zhu18, He20}. We employ the bimodal ``dichotomy" heuristic hereafter in this work.

Single-transiting planetary systems contain one known planet, which transits in front of its host star. However, it is possible that these systems contain $N \geq 1$ undiscovered planets which may have longer orbital periods or large mutual inclinations \citep{Ballard16}. Additionally, \textit{Kepler} has discovered hundreds of multi-transiting planetary systems which are dynamically packed \ben{\citep{Fabrycky14, Winn15}}. Here, we define compact multiple planetary systems as containing multiple transiting planets in a dynamically packed arrangement. Planets in compact multiple systems have a propensity to have similar planetary parameters, such as similar sizes, masses, and orbital period ratios \ben{\citep{Millholland17, Weiss18}}. This ``peas in a pod" observation appears to break down for the gas giant planets (\ben{$\ge \ \sim$100 \mearth \ and $\ge \ \sim$10 \rearth}; \cite{Otegi22}).

Aside from the ``$Kepler$ dichotomy" and orbital architecture of planetary systems, the planet occurrence rate also appears to be correlated with bulk metallicity. The probability of a star hosting a giant planet was found to increase significantly for stars with metallicities greater than that of the Sun \citep{Fischer05}. More recently, statistical evidence has been presented showing that compact multiple planetary systems either form more readily or have a higher probability of survival to field age around more metal-poor stars than metal-rich stars \citep{Brewer18, Anderson21}. The key to these observations is the construction of a both a uniform planetary sample as well as uniformly derived stellar parameters (including metallicity). One of the most significant results enabled through the derivation of uniform stellar samples is the discovery of the Fulton Gap \citep{Fulton17}. The Fulton Gap is a feature of the planetary radius distribution at roughly 1.7 \rearth \ which defines a gap between two bimodal peaks in the distribution (possibly indicating a divide between rocky planets and water worlds (\citet{Lozovsky18}; although understanding the origin of this feature remains a subject of ongoing research). The ``peas in a pod" correlation becomes stronger when considering planets on same sides of the Fulton Gap \citep{Millholland21}.

Recently, \cite{RodriguezMartinez23} performed a  study using a diverse set of 70 planets with radial velocity detection around M dwarfs \ben{to measure the difference between the density distributions of compact multiples and single-transiting planetary systems}. Although this study found a tentative difference between theses samples for planets with radii greater than 4 \rearth, it was inconclusive for planets less than 4 \rearth \ \citep{RodriguezMartinez23}. In this paper we investigate the radius distribution of single-transiting and compact multiple planets orbiting late K and M dwarfs with a particular focus on the small planet population in and around the Fulton Gap and find significant differences between these two populations. 

This paper is organized as follows. In Section \ref{sec:DataObs} we present our stellar sample and derived uniform stellar parameters. We employ these uniform stellar parameters to calculate radius values for planets around these host stars. In Section \ref{sec:analysis} we test the uniformity of our stellar sample before investigating various subsamples of our planetary sample to determine if there is a statistically significant difference in the radius distribution of compact multiple planets and single-transiting planets. Specifically, we perform two-sample Kolmogorov-Smirnov (K-S) tests, Mann-Whitney U tests, and Anderson Darling k-sampling on our compact multiple- and single-transiting planet samples. We find that compact multiple planets are, on average, larger than their single-transiting planet counterparts and describe the statistical significance of our finding. In Section \ref{sec:discussion} we investigate possible origins of this effect including whether this effect is confined to systems in  resonant orbits and possible effects of extended atmospheres from secondary atmosphere degassing. In Section \ref{sec:conclusion} we summarize our procedures and findings.

\section{Data and Observations}
\label{sec:DataObs}
\ben{In this work we used data from the Kepler Input Catalog (KIC) \citep{Brown11} and the NASA Exoplanet Archive \citep{Akeson13}. \footnote{\ben{All the {\it KIC} data can be found in doi: \dataset[10.17909/T9059R]{http://dx.doi.org/10.17909/T9059R} \citep{doi.org/10.17909/t9059r} and the {\it Exoplanet Archive} data can be found in IPAC: doi:\dataset[10.26133/NEA5]{http://dx.doi.org/10.26133/NEA5} \citep{koidr25}.}}} We base our sample on that of \citet{Anderson21}, which we have reconstructed here. This subsample of stars contains 7146 M and late K dwarfs \ben{with effective temperatures less than 4500 K} selected from the KIC \citep{Brown11}. Although there is some red giant contamination in the original sample of \citet{Anderson21}, here we are interested in the detected planet sample, which is unaffected by this potential bias (as $Kepler$ is not sensitive to small planets around giant stars).  Of the 7146 stars in the \cite{Anderson21} sample, 207 are $Kepler$ objects of interest likely to contain transiting planetary systems. We investigated 203 of these planetary systems which we further subdivide into single- and multi-transiting planetary systems. The remaining four planetary systems were removed from our sample because they have been found to be false positives as reported in the Exoplanet Archive: KIC9142714, KIC4063943, KIC9224514, and KIC6620003 \citep{Akeson13}. 

\begin{figure}[h]
    \centering
    \includegraphics[width=0.5\linewidth]{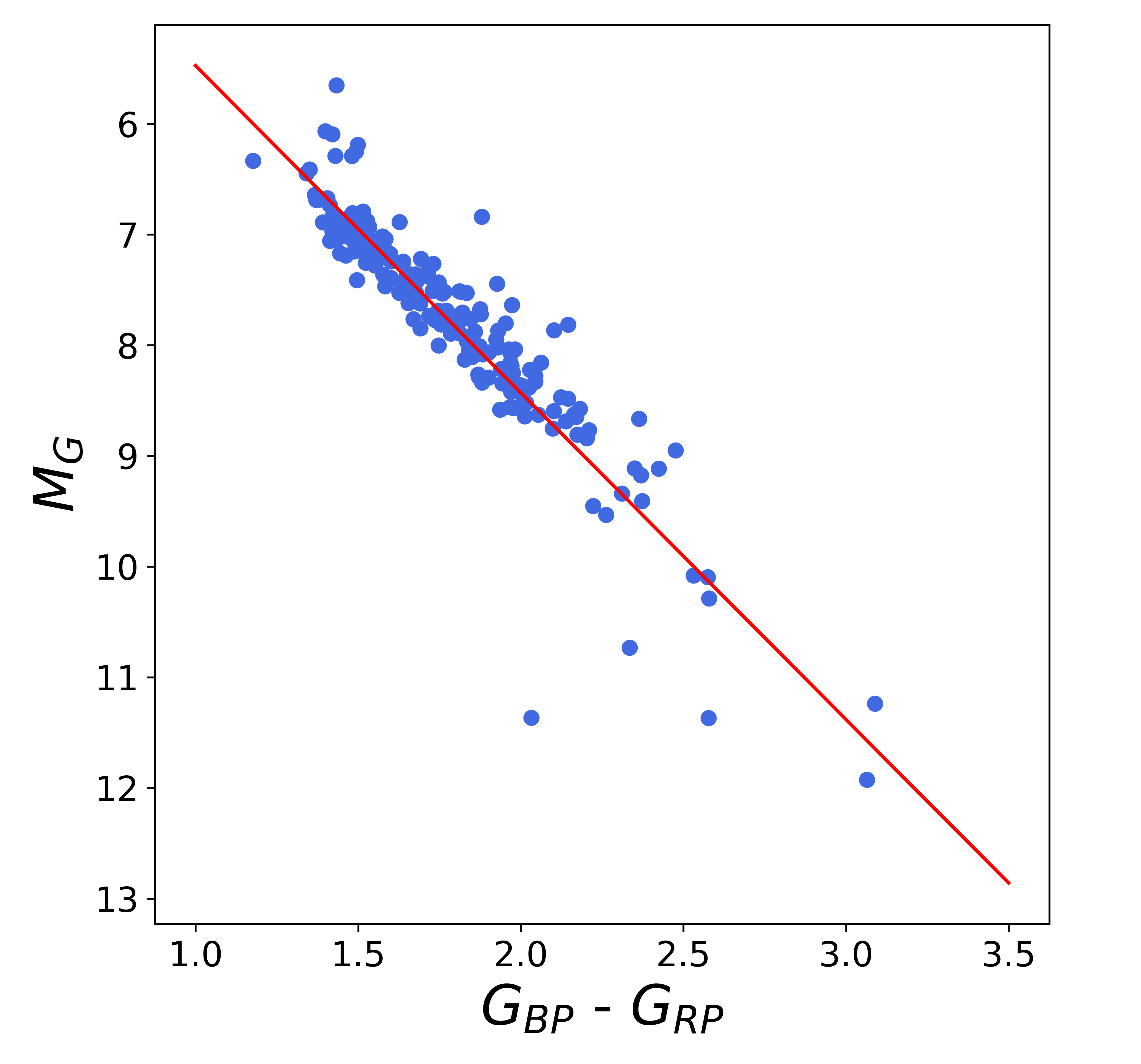}
    \caption{Color-magnitude diagram of host stars in \textit{Gaia} $G_G$ vs $G_{BP}-G_{RP}$. \cite{Anderson21} found that the linear fit to these stars represents a boundary of average metallicity for main-sequence stars, with most compact multiple host stars occupying the left of the linear trend line in the metal-poor regime.}
    \label{fig:color_color}
\end{figure}

We have compiled a list of stellar magnitudes for our  203 planetary host stars through querying both the KIC \citep{Brown11} and \textit{Gaia} Data Release 3 (DR3; \citet{Gaia2016b, Gaia2022k}) to generate Table \ref{tab:star_table}. Figure \ref{fig:color_color} shows a color-magnitude diagram of the host stars. Our statistical analysis will rely on obtaining uniform stellar (and therefore uniform planetary) parameters, which has only recently become feasible with these large uniform data sets. These uniformly derived parameters will be largely insulated from the systematic errors associated with combining data from multiple surveys together.

\begin{table}[ht]
\caption{Stellar Parameters} 
\centering 
\begin{tabular}{c c c} 
\hline\hline 
Column Number & Units & Description \\ [0.5ex] 
\hline 
1 & ... & \textit{Gaia} Source ID \\ 
2 & deg & R.A. \\
3 & mas & R.A. Error \\
4 & deg & decl. \\
5 & mas & decl. Error \\
6-7 & mas & Parallax, \ Parallax Error \\
8-9 & $e^- /s$ & $G_G$ Flux, \ $G_G$ Flux Error \\
10 & mag & $G_G$ \\
11-12 & $e^- /s$ & $G_{BP}$ Flux, \ $G_{BP}$ Flux Error \\
13 & mag & $G_{BP}$ \\
14-15 & $e^- /s$ & $G_{RP}$ Flux, \ $G_{RP}$ Flux Error \\
16 & mag & $G_{RP}$ \\
17-19 & K & $T_{eff}$, \ $T_{eff}$ Error \\
20-21 & mag & $J$, \ $J$ Error\\
22-23 & mag & $H$, \ $H$ Error \\
24-25 & mag & $Ks$, \ $Ks$ Error \\
26-27 & mag & $g$, \ $g$ Error \\
28-29 & mag & $r$, \ $r$ Error \\
30-31 & mag & $i$, \ $i$ Error \\
32 & ... & \textit{Kepler} ID \\ 
33-35 & \rsun & $R_{\star}$, \ $R_{\star}$ Error \\
36-38 & \msun & $M_{\star}$, \ $M_{\star}$ Error \\ [1ex] 
\hline 
\end{tabular}
\begin{tablenotes}
      \small
      \centering\item (This table is available in its entirety in machine-readable form.)
    \end{tablenotes}
\label{tab:star_table} 
\end{table}

Stellar parameters for 194 of the 203 host stars were pulled from \cite{Berger23}. These 194 planetary systems consist of 141 single-transiting systems and 53 compact multiple systems. \ben{$\sim$ 71 of these stars are M dwarfs and $\sim$ 123 are late K dwarfs.} We define two subset samples in our stellar population coming from \cite{Berger23}. The first subset contains 170 stars who also have transit depth values reported from the NASA Exoplanet Archive \citep{Akeson13}. For this collection of stars, we took stellar radius values from \cite{Berger23} and transit depths and orbital periods from the Exoplanet Archive to calculate planetary radius values. The first subset of stars contains 117 single-transiting planetary systems and 53 multi-transiting systems. The second subset of 24 stars did not have transit depth values reported from the Archive. For these systems, planetary radius values were derived using the $R_{p} / R_{\star}$ and stellar radius values from \cite{Berger23}. Orbital periods for this second subset were pulled from \cite{Berger23}. The second subset was made up of only single-transiting systems. Therefore, for these systems, our parameters are largely derived in a uniform manner as reported by \citet{Berger23}. 

\cite{Berger23} did not present stellar parameters for the nine remaining host stars. Stellar radius values  for the missing nine host stars were obtained from \cite{Latham05}. Similarly to the \cite{Berger23} stars, we define two subset samples in the \cite{Latham05} stars. The first subset contains six stars for which planetary radius values were derived from \cite{Latham05} stellar radius values and Exoplanet Archive transit depths. These six systems contain five single-transiting systems and one compact multiple. The second subset contains three stars which did not have reported transit depth values on the Exoplanet Archive. For these three systems, planetary radius values were calculated using $R_{p} / R_{\star}$ ratios reported on the Exoplanet Archive and \cite{Latham05} stellar radius values. These three systems were all single-transiting systems. Orbital periods for all planets around the nine host stars were pulled from the Exoplanet Archive. Therefore, this minor subsample itself is derived in a uniform manner in order to limit the effect of possible systematics associated with the different methodology and data for these nine of 203 host stars. 

\begin{table}[ht]
\caption{Categorization of Planetary Systems Analyzed} 
\centering 
\begin{tabular}{c c c} 
\hline\hline 
Type of Planetary System & Number of Planetary Systems & Number of Planets \\ [0.5ex] 
\hline 
Single-transiting System & 149 & 149 \\ 
Compact Multiple & 54 & 141 \\
Total & 203 & 290 \\ [1ex] 
\hline 
\end{tabular}
\label{table:planetPop} 
\end{table}

\begin{figure}[ht]
    \centering
    \includegraphics[width=0.5\linewidth]{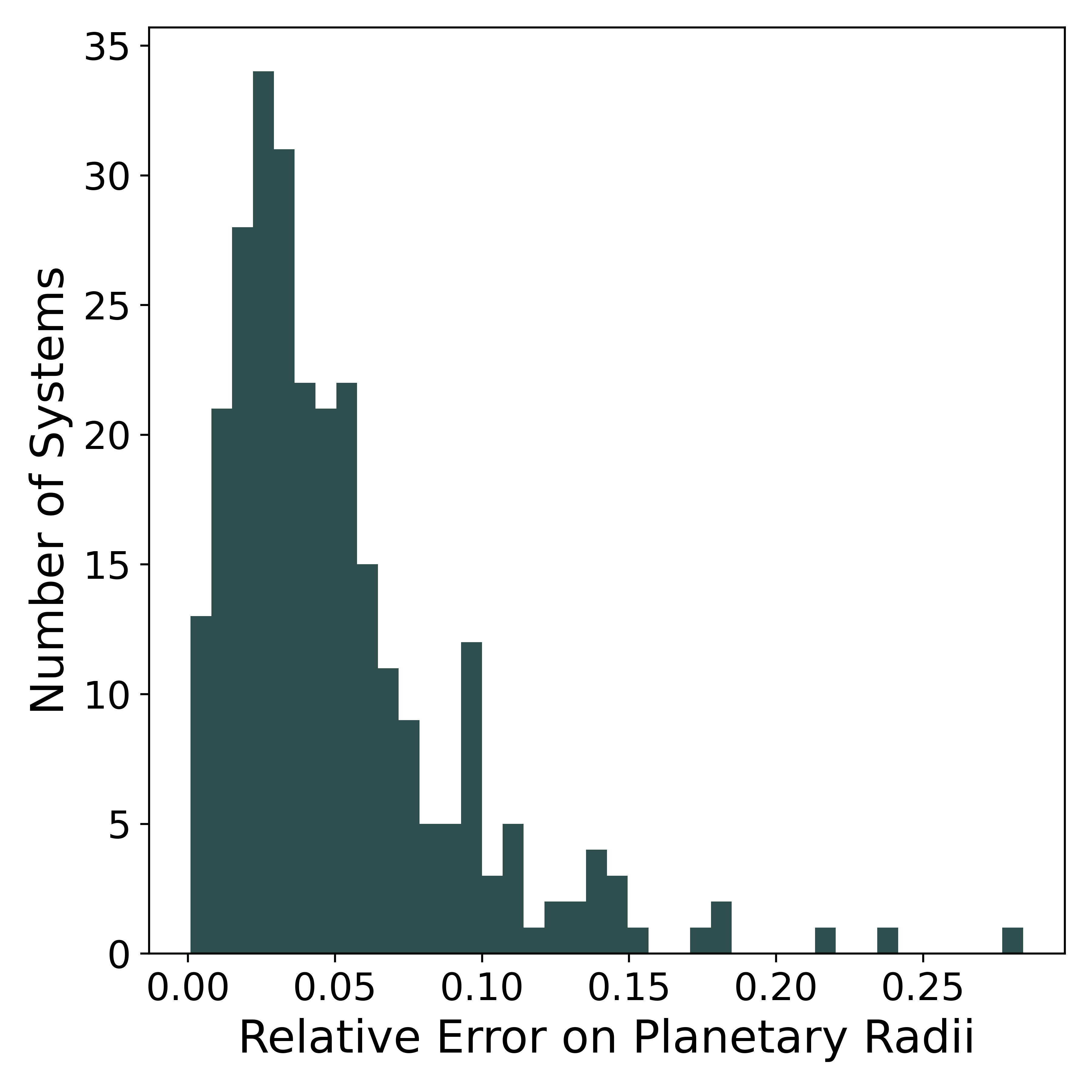}
    \caption{Histogram of the relative percent errors ($\sigma_{R_p} / R_{p}$) on derived planetary radii. Relative percent errors on planetary radius values for 263 planets were propagated from errors on the transit depth such that $\sigma_{R_p} = \frac{\sigma_{\delta}}{\delta}$. For the 27 remaining planets in our sample relative percent errors were calculated from $\sigma_{R_p} = \frac{\delta_{R_p / R_\star}}{R_p / R_\star}$. Our relative errors on planetary radii do not factor in stellar uncertainties since our planetary sample is built from a uniform stellar sample.
    }
    \label{fig:rel_error_hist}
\end{figure}

The entirety of our planet population is broken down demographically in Table \ref{table:planetPop}. \ben{Radius values of our planet population ranged from 0.381 \rearth \ to 29.6 \rearth.} Relative percent errors ($\sigma_{R_p} / R_{p}$) on the radius measurements for each planet were propagated from errors on the transit depth or $R_p / R_{\star}$ such that $\sigma_{R_p} = \frac{\sigma_{\delta}}{\delta}$ or $\sigma_{R_p} = \frac{\delta_{R_p / R_\star}}{R_p / R_\star}$, respectively. The methodology for each planet followed depending on if $R_p$ was derived from a transit depth or a $R_p / R_\star$ ratio. Two-hundred sixty-three radius values were derived using reported transit depths $ (R_p / R_\star)^{2} $. Twenty-seven planets were calculated using reported $R_p / R_\star$ ratios. \ben{The three largest planets in our sample, $R_p = \{29.6 \ \text{\rearth}, 27.4 \ \text{\rearth}, 25.2 \ \text{\rearth}\}$, have enormous relative errors on their radius values. These planets are cut from our sample when we perform a quality cut in our analysis. A more representative range of the radius regime considered in our study is 0.381 \rearth \ to 15.1 \rearth.} Figure \ref{fig:rel_error_hist} showcases the distribution of $\sigma_{R_p} / R_{p}$ for planets \ben{with radius values between 0.381 \rearth \ and 15.1 \rearth.}

We establish a quality cut at the 95\% percentile of our sample. We make this choice because we are interested in variations between the radius distributions of the two samples (singles and compact multiples), and therefore if the uncertainty on planetary radius is too high, that planet ceases to offer relevant distinguishing statistical power. The remaining 274 planets in our sample all have radius determinations with a relative error  of less than 23\% (without uncertainties in the stellar modes). 
We show the population breakdown in Table \ref{table:qualityPlanetPop}. Table \ref{tab:planet_table} presents the list of planet parameters considered in our study for all 290 planets.

\begin{table}[ht]
\caption{Categorization of Planetary Systems Analyzed Post-Quality Cut} 
\centering 
\begin{tabular}{c c c} 
\hline\hline 
Type of Planetary System & Number of Planetary Systems & Number of Planets \\ [0.5ex] 
\hline 
Single-transiting System & 135 & 135 \\ 
Compact Multiple & 54 & 139 \\
Total & 189 & 274 \\ [1ex] 
\hline 
\end{tabular}
\label{table:qualityPlanetPop} 
\end{table}

\begin{table}[ht]
    \centering
    \caption{Planet Parameters}
    \begin{tabular}{ccccc} 
    \hline
    \hline
         Host KIC ID&  PS Type&   $R_{p}$ [\rearth]&  $\sigma_{R_p} / R_{p}$& P [days]\\ 
         \hline
         kic6425957 & m & \ben{1.589} & \ben{0.012} & \ben{2.756}\\ 
         kic6425957 & m & \ben{1.834} & \ben{0.018} & \ben{20.307} \\ 
         kic5871985 & s & \ben{3.314} & \ben{0.031} & \ben{213.259}\\
         kic8509442 & s & \ben{2.445} & \ben{0.068} & \ben{82.660} \\ 
         kic11754553 & m & \ben{2.319} & \ben{0.037} & \ben{16.385} \\ 
    \hline
    \end{tabular}
    \begin{tablenotes}
      \small
      \centering\item (This table is available in its entirety in machine-readable form.)
    \end{tablenotes}
    \label{tab:planet_table}
\end{table}

\section{Analysis and Results}
\label{sec:analysis}

Since we now have uniformly derived physical parameters, we now turn to examining the radius distribution of the single-transiting planet and the compact multiple-transiting planet systems.

\subsection{Stellar Sample}
\begin{figure}[ht]
    \centering
    \includegraphics[width=0.75\linewidth]{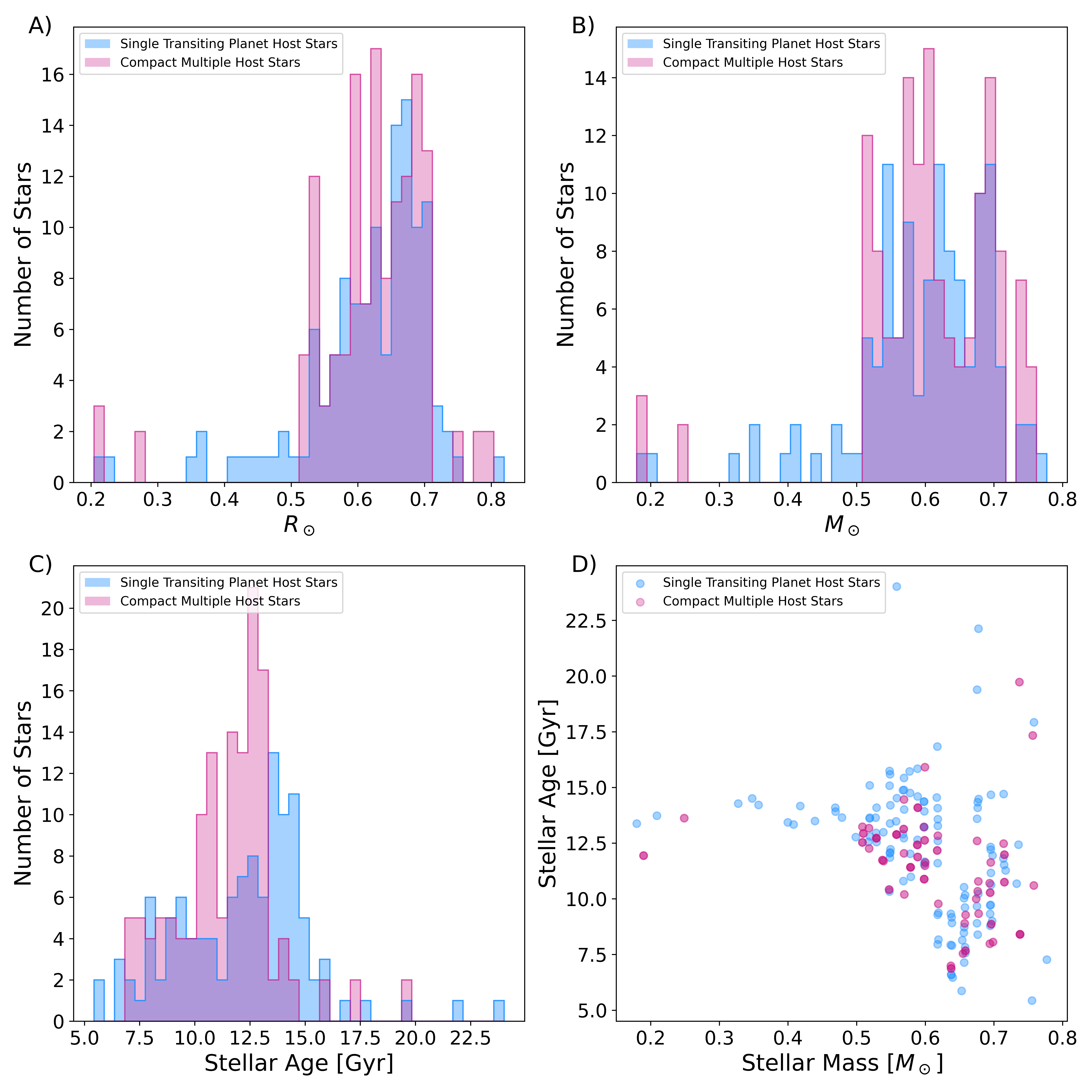}
    \caption{\textbf{A)} Overlaid histograms of stellar radii of single transiting planet host stars (blue) and compact multiple system host stars (pink). \textbf{B)} Overlaid histograms of stellar mass of single transiting planet host stars (blue) and compact multiple system host stars (pink).  \textbf{C)} Overlaid histograms of stellar age of single transiting planet host stars (blue) and compact multiple system host stars (pink). The uncertainty on the stellar ages used are immense. \textbf{D)} Stellar age vs stellar mass for host stars of single transiting planetary (blue) and compact multiple systems (pink). Most of our stellar sample host stars have stellar radius values between 0.5 \rsun \ and 0.7 \rsun \ and stellar mass values between 0.5 \msun \ and 0.7 \msun \ with a tail towards smaller masses and radii. The single-transiting planetary systems and the compact multiples do not have significantly different stellar parameters and derived differences in the planetary samples are unlikely to be a consequence of their host star population. } 
    \label{fig:stars_hist}
\end{figure}

Firstly, we investigate whether any potential differences between our planetary sample could be the consequence of statistical difference in the stellar sample itself rather than the physical characteristics of the planets. We divide our stellar population into two groups based off of whether a particular host star harbors only one known planet or multiple planets. We plot histograms of the stellar radii and stellar masses in Figure \ref{fig:stars_hist}. The bulk of both the single-transiting planet host stars and compact multiple host stars samples have stellar radius values between 0.5 \rsun \ and 0.7 \rsun \ and stellar mass values between 0.5 \msun \ and 0.7 \msun \ with a tail towards smaller masses and radii. We applied a two-sample K-S test, Mann-Whitney U test, and Anderson-Darling k-sampling test on the stellar radius and mass samples shown in Figure \ref{fig:stars_hist}. The results of these statistical tests are displayed in Table \ref{tab:stellar_radii_stat_tests} and Table \ref{tab:stellar_mass_stat_tests} for stellar radii and stellar masses, respectively. The lack of significant deviance between the two samples in the statistical tests suggests that the stellar populations between the single-transiting planetary systems and the compact multiples are not significantly different and derived differences in the planetary samples are unlikely to be a consequence of their host star population. Taking this into consideration with our temperature cutoff of 4500 K allows us to make population-level statistical arguments for the planets orbiting the stars in our sample despite some of the host stars being split between the late K and M dwarf spectral classes.

\begin{table}[ht]
    \centering
    \caption{Statistical Tests for Stellar Radii}
    \begin{tabular}{ccccc} 
    \hline
    \hline
         Statistical Test&  p-value & Statistic\\ 
         \hline
         Two-sample K-S Test &  0.7651 & 0.0793\\ 
         Mann-Whitney U Test& 0.9890 & 8592.0 \\ 
         Anderson-Darling K-sampling Test & 0.250 & -0.742\\ 
    \hline
    \end{tabular}
    \label{tab:stellar_radii_stat_tests}
\end{table}

\begin{table}[ht]
    \centering
    \caption{Statistical Tests for Stellar Masses}
    \begin{tabular}{ccccc} 
    \hline
    \hline
         Statistical Test&  p-value & Statistic\\ 
         \hline
         Two-sample K-S Test &  0.7513 & 0.0815\\ 
         Mann-Whitney U Test& 0.4121 & 7647.0 \\ 
         Anderson-Darling K-sampling Test & 0.250 & -0.085\\ 
    \hline
    \end{tabular}
    \label{tab:stellar_mass_stat_tests}
\end{table}

We also attempted to investigate the age of our sample. Stellar ages are important to consider since they provide context as to the maturity of a planetary system and the timescale for which certain planetary processes may have occurred. If compact multiples self-disrupt on some timescale, we might expect single planetary systems to be the remnants of these disrupted systems and would measure a difference in the age distributions in the host stars. We acquired stellar ages as reported from \cite{Berger23} for 117 of the single-transiting planet host stars and 53 of the compact multiple host stars in the sample. Figure \ref{fig:stars_hist}C displays a histogram of stellar age between the two subsamples. Figure \ref{fig:stars_hist}D shows a scatter of the masses and ages of the host stars in our sample. We do not observe any significant difference between the ages of these stars, although we note that the uncertainty and difficulty in deriving stellar ages makes it difficult to draw any robust conclusions about the average ages of our two subsamples. 

\subsection{Full Planetary Sample}
\begin{figure}[ht]
    \centering
    \includegraphics[width=1\linewidth]{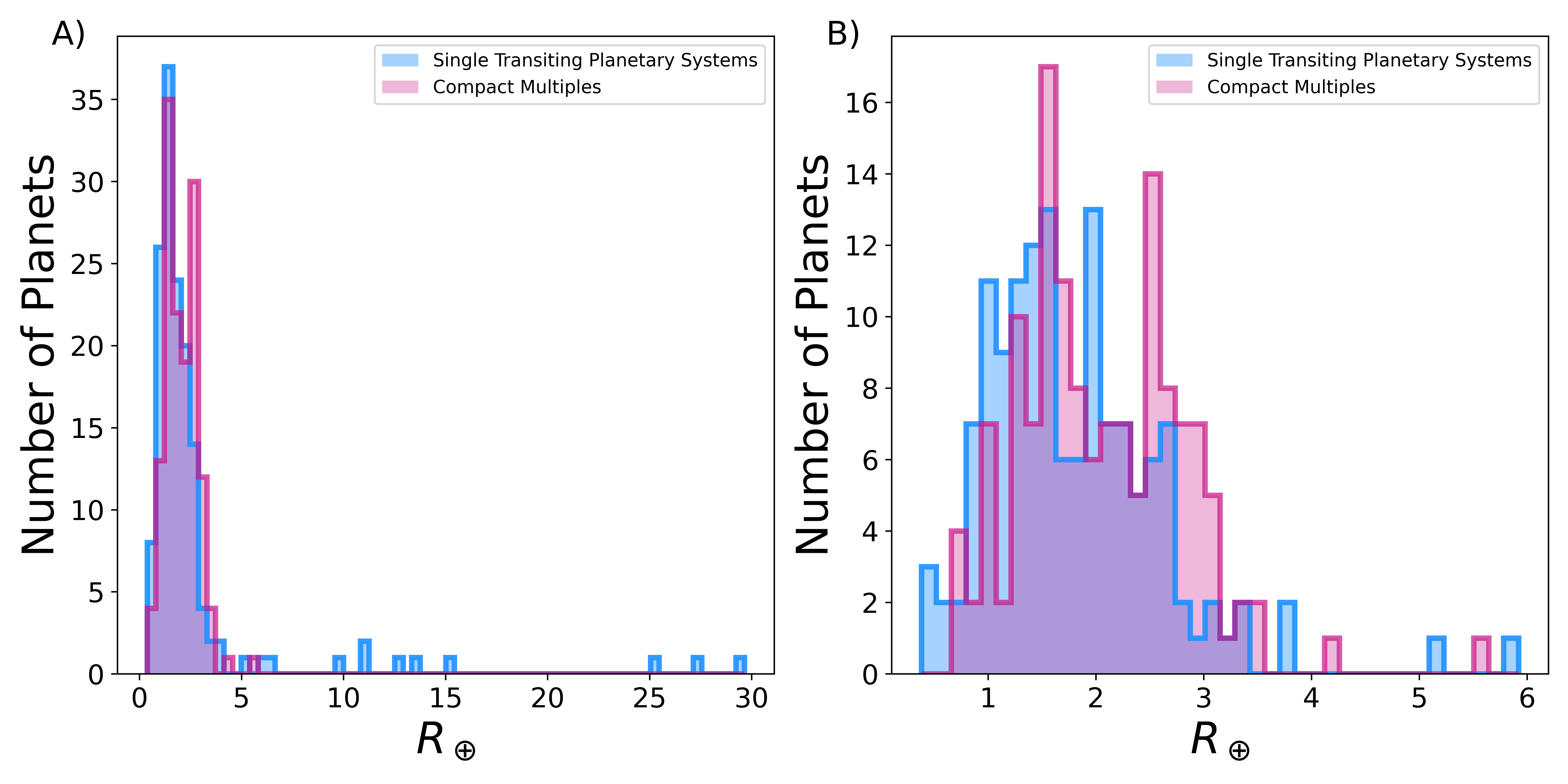}
    \caption{\textbf{A)} Overlaid histograms of planetary radius distributions for planets in single transiting systems (blue) and compact multiple planets (pink). \textbf{B)} Overlaid histograms of planetary radius distributions for planets in single transiting systems (blue) and compact multiple planets (pink) with radius values less than 6 \rearth. We performed a two-sample K-S test, a Mann-Whitney U test, or Anderson-Darling k-sampling test for both the full planetary and consideration of only planets $\leq$ 6 \rearth. In Tables \ref{tab:planet_all_stat_tests} and \ref{tab:planet_under6_stat_tests} we present the results of the statistical tests, finding the radius distributions of single-transiting planets and planets in compact multiples to be significantly different in all statistical tests. Importantly, the test results become more significant (p $<$ 0.0009) when only accounting for planets $\leq$ 6 \rearth. The inclusion or exclusion of gas giant planets is not significant when assessing the radius distributions of these two samples.}
    \label{fig:planet_hist}
\end{figure}

Finally, we turn to the statistical differences between the derived planetary parameters themselves. We begin our analysis of our planetary sample by looking at all 290 planets regardless of their $\sigma_{R_p} / R_{p}$ value. We divide all planets into two groups; the first group represents our single-transiting planetary systems where there is only one known planet in the planetary system and the second group represents planets which are in planetary systems that have multiple known planets transiting around their host star. 
We display the overlaid histograms of planets in single-transiting planetary systems and compact multiple systems in Figure \ref{fig:planet_hist}A. Figure \ref{fig:planet_hist}B zooms in on the planets with $R_{p} \leq$ 6 \rearth.
\ben{The single-transiting systems have a median radius value of 1.688 \rearth \ and 1.626 \rearth \ in panels A and B of Figure \ref{fig:planet_hist} respectively. The median compact multiple planetary radius value is 1.940 \rearth. In both panels of Figure \ref{fig:planet_hist} the median of the compact multiple planets is larger than their single-transiting counterparts.}

\begin{table}[ht]
    \centering
    \caption{Statistical Tests for Full Planetary Sample}
    \begin{tabular}{ccccc} 
    \hline
    \hline
         Statistical Test&  p-value & Statistic\\ 
         \hline
         Two-sample K-S Test &  0.0261 & 0.1698\\ 
         Mann-Whitney U Test& 0.0295 & 8950.0 \\ 
         Anderson-Darling K-sampling Test & 0.009 & 3.865\\ 
    \hline
    \end{tabular}
    \label{tab:planet_all_stat_tests}
\end{table}

\begin{table}[ht]
    \centering
    \caption{Statistical Tests for Planets Under 6 \rearth}
    \begin{tabular}{ccccc} 
    \hline
    \hline
         Statistical Test&  p-value & Statistic\\ 
         \hline
         Two-sample K-S Test &  0.0071 & 0.1973\\ 
         Mann-Whitney U Test& 0.0009 & 7540.0 \\ 
         Anderson-Darling K-sampling Test & 0.001 & 6.561\\ 
    \hline
    \end{tabular}
    \label{tab:planet_under6_stat_tests}
\end{table}

We note there are no planets in the compact multiple sample which have a $R_{p} >$ 6 \rearth. Therefore, we conduct an additional test between the single-transiting planetary systems and compact multiple-transiting systems by eliminating these larger planets from the single-transiting planetary systems and repeating our statistical tests.

Tables \ref{tab:planet_all_stat_tests} and \ref{tab:planet_under6_stat_tests} showcase the statistical test results of a two-sample K-S test, Mann-Whitney U test, and Anderson-Darling k-sampling test for both of these cases. We find the radius distributions of single-transiting planets and planets in compact multiples to be significantly different in all statistical tests for both the full planetary sample and consideration of only planets $\leq$ 6 \rearth. Importantly, the test results become more significant when only accounting for planets $\leq$ 6 \rearth, suggesting that the inclusion or exclusion of gas giant planets is not significant when assessing the radius distributions of these two samples. \ben{Even though including the single gas giant planets in our single-transiting planet distribution slightly lowers our statistical significance (as the compact multiple planets are larger, so adding large planets to the single population pulls the median), we suggest this is not a useful comparison. By eliminating the gas giants, we are more directly comparing rocky planets in single-planet configurations with rocky planets in multi-planet configurations.} Here we considered p-values less than 0.05 to denote a significant deviation between our samples.

\ben{From the Mann-Whitney U test, we find a p-value of 0.0009 supporting that planets in compact multiple systems are larger, on average, when compared with their single-transiting counterparts. We adopt this value since the Mann-Whitney U test} is primarily sensitive to differences in the averages (central tendency) of two distributions and does not require Gaussianity like other tests (such as a t-test), which may be relevant since we are probing the Fulton Gap regime. 
\ben{We find that the significance of our p-value is preserved} when performing a jackknife resampling sensitivity test with a 10\% elimination factor, suggesting that this result is not driven by a handful of outliers in the sample. On each iteration of our jackknife test, a random 10\% of planets in both the compact multiples and single-transiting planets are removed from their respective samples before the Mann-Whitney U test is reapplied to the reduced samples. The p-value obtained from the Mann-Whitney U test rarely loses significance. Therefore, we do not believe our result to be due to variations within the small sample number we investigated.

\subsubsection{Uniform Planetary Subsample}
Although the vast majority (all but nine) of our systems have parameters derived from the study of \cite{Berger23}, here we perform a test to ensure that this small minority of systems is not responsible for the significance of our results. When we consider only the planets that had radius values derived from \cite{Berger23} we are left with the 194 planetary systems depicted in Table \ref{table:berger_planets}. We note again that the 280 planets in this subsample all had planetary radius values derived in a consistent manner from stellar parameters that were uniformly determined by \cite{Berger23}.

\begin{table}[ht]
\caption{Categorization of Uniform Planetary Systems Analyzed from \cite{Berger23}} 
\centering 
\begin{tabular}{c c c} 
\hline\hline 
Type of Planetary System & Number of Planetary Systems & Number of Planets \\ [0.5ex] 
\hline 
Single-Transiting System & 141 & 141 \\ 
Compact Multiple & 53 & 139 \\
Total & 194 & 280 \\ [1ex] 
\hline 
\end{tabular}
\label{table:berger_planets} 
\end{table}

This Berger-subgroup \citep{Berger23} of planets present a fully uniform sample of planets for which we performed additional statistical tests. We repeated a two-sample K-S test, Mann-Whitney U test, and Anderson-Darling k-sampling test for the same divisions of single-transiting planets compared to compact multiples as in the case of the full planetary sample. The results of these statistical tests are shown in Tables \ref{tab:berger_planet_all_stat_tests} and \ref{tab:berger_planet_under6_stat_tests}. The statistical tests still indicate that compact multiple planets come from a different \ben{underlying} population than the single-transiting planets in both considerations of the single-transiting planets. The significant results also continue to hold under our jackknife test. Therefore, we conclude that the variations between the compact multiple planetary systems radius distribution and the single transiting planet radius distribution cannot be ascribed to differences in the stellar sample such as mass, radius, or age (although the stellar ages of our sample are very poorly constrained) nor any underlying non-uniformity in the derivation of the parameters of the planets themselves. The most likely explanation is that the observed differences are real and inherent to the physical properties of the planetary systems themselves. 

\begin{table}[ht]
    \centering
    \caption{Statistical Tests for Full Uniform Planetary Subsample with Stellar Parameters from \cite{Berger23}}
    \begin{tabular}{ccccc} 
    \hline
    \hline
         Statistical Test&  p-value & Statistic\\ 
         \hline
         Two-sample K-S Test &  0.0474 & 0.1601\\ 
         Mann-Whitney U Test& 0.0444 & 8437.0 \\ 
         Anderson-Darling K-sampling Test & 0.013 & 3.475\\ 
    \hline
    \end{tabular}
    \label{tab:berger_planet_all_stat_tests}
\end{table}

\begin{table}[ht]
    \centering
    \caption{Statistical Tests for Uniform Planets with Stellar Parameters from \cite{Berger23} Under 6 \rearth}
    \begin{tabular}{ccccc} 
    \hline
    \hline
         Statistical Test&  p-value & Statistic\\ 
         \hline
         Two-sample K-S Test &  0.0136 & 0.1888\\ 
         Mann-Whitney U Test& 0.0013 & 7047.0 \\ 
         Anderson-Darling K-sampling Test & 0.002 & 6.024\\ 
    \hline
    \end{tabular}
    \label{tab:berger_planet_under6_stat_tests}
\end{table}

\subsection{Planetary Sample Post-Quality Cut ($\sigma_{Rp} / R_p \leq 0.2299$)}

\label{sec:3.3}

\begin{figure}[ht]
    \centering
    \includegraphics[width=1\linewidth]{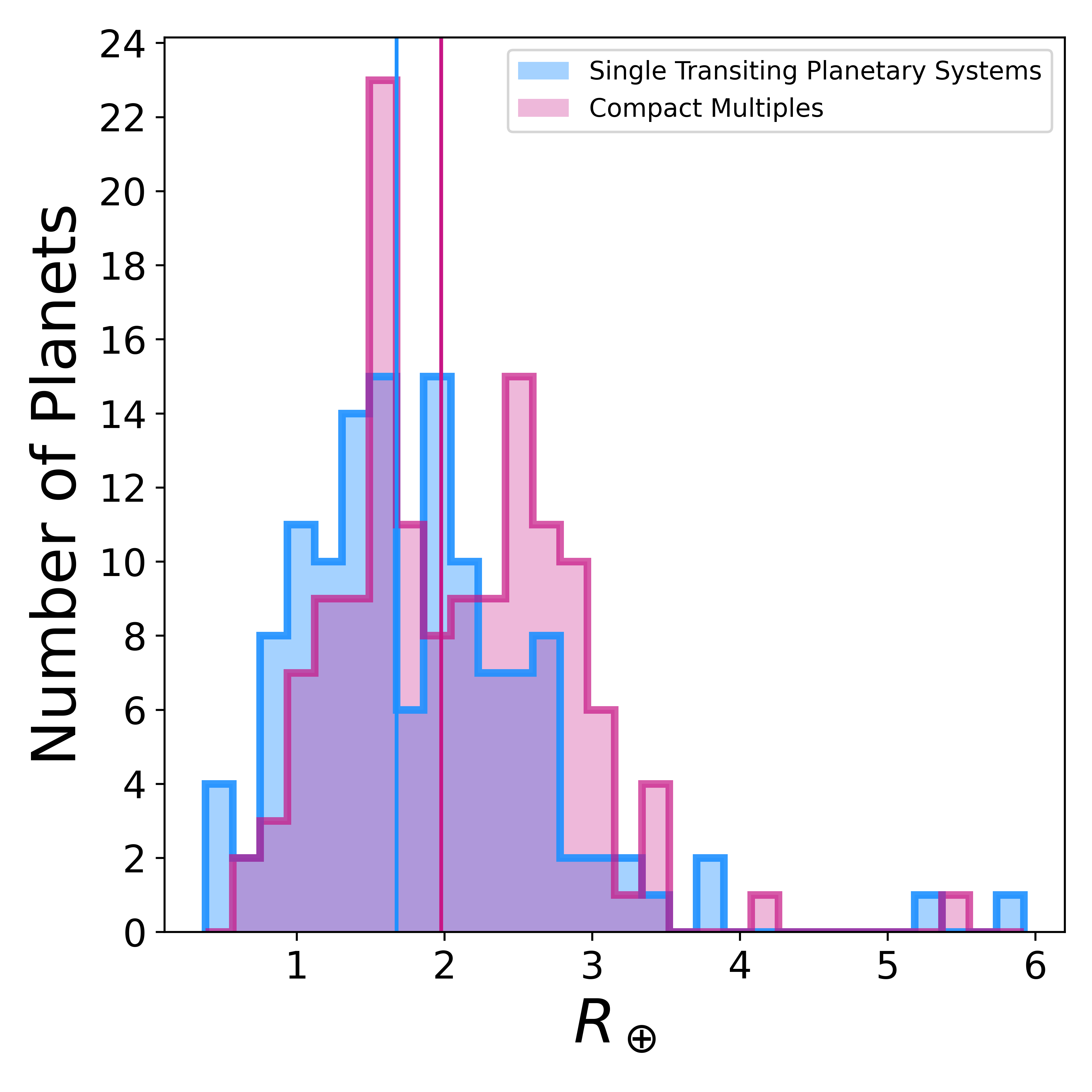}
    \caption{Overlaid histograms of the planetary radius distributions for planets in single-transiting systems with radius values less than 6 \rearth \ (blue) and compact multiple planets (pink) after performing a 95th percentile quality cut on both samples.  The median of each sample is plotted as a vertical line of the same color. The median lines are off-shifted by 0.301 \rearth. Our quality cut threw out 15 planets from our study which had $\sigma_{Rp} / R_p > 0.2299$. The two-sample K-S test, Mann-Whitney U test, and Anderson-Darling k-sampling test were reapplied to the planets surviving the quality cut. Tables \ref{tab:quality_planets_all_stat_tests} and \ref{tab:quality_planets_under6_stat_tests} display the significance of our statistical tests which endured our jackknife resampling.}
    \label{fig:quality_planet_hist}
\end{figure}

To test if our significant statistical result was a consequence of the uncertainty associated on our derived planetary radius values, we accepted a 95th percentile quality cut in our sample based on the derived $\sigma_{Rp} / R_p$ plotted in Figure \ref{fig:rel_error_hist}. This cut threw out 15 planets from our study which had $\sigma_{Rp} / R_p > 0.2299$. The demographics of the remaining planets are outlined in Table \ref{table:qualityPlanetPop}. In Figure \ref{fig:quality_planet_hist} we plot the planetary radius values that survived the quality cut and have $R_p \leq 6$ \rearth. There are 128 single-transiting systems with planets that have $R_p \leq 6$ \rearth. Additionally, the median radius value for each planetary sample is plotted as a vertical line in Figure \ref{fig:quality_planet_hist}. The two-sample K-S test, Mann-Whitney U test, and Anderson-Darling k-sampling test were reapplied to the planets surviving the quality cut. The significance of these tests, displayed in Tables \ref{tab:quality_planets_all_stat_tests} and \ref{tab:quality_planets_under6_stat_tests}, also endured our jackknife resampling.

\begin{table}[ht]
    \centering
    \caption{Statistical Tests for All Quality Planets ($\sigma_{Rp} / R_p \leq 0.2299$)}
    \begin{tabular}{ccccc} 
    \hline
    \hline
         Statistical Test&  p-value & Statistic\\ 
         \hline
         Two-sample K-S Test &  0.0481 & 0.1615\\ 
         Mann-Whitney U Test& 0.0315 & 7972.0 \\ 
         Anderson-Darling K-sampling Test & 0.013 & 3.418\\ 
    \hline
    \end{tabular}
    \label{tab:quality_planets_all_stat_tests}
\end{table}

\begin{table}[ht]
    \centering
    \caption{Statistical Tests for Quality Planets ($\sigma_{Rp} / R_p \leq 0.2299$) Under 6 \rearth}
    \begin{tabular}{ccccc} 
    \hline
    \hline
         Statistical Test&  p-value & Statistic\\ 
         \hline
         Two-sample K-S Test &  0.0210 & 0.1814\\ 
         Mann-Whitney U Test& 0.0026 & 6999.0 \\ 
         Anderson-Darling K-sampling Test & 0.003 & 5.296\\ 
    \hline
    \end{tabular}
    \label{tab:quality_planets_under6_stat_tests}
\end{table}

\section{Discussion}
\label{sec:discussion}


We have identified a small, but statistically significant, offset between the radii of planets in multi-transiting systems, and planets in single-transiting systems. We quantify the degree of significance with several metrics and also by considering subsamples of our planet population. We express particular interest in looking at the subgroup of single-transiting planets with $R_p \leq 6$ \rearth \ since this is the size range which all of our compact multiple planets fall into. In consideration of the planets remaining after the quality cut, the difference between the median radius values of planets in single-transiting and multi-transiting planetary systems is 0.301 \rearth. This is a substantial size difference when the median sized planet in the single-transiting systems is 1.676 \rearth, and the median sized planet in our compact multiple sample is 1.976 \rearth. To test the validity of such a sizable shift between the two populations, the single-transiting systems were artificially inflated to the points where the Mann-Whitney U test would no longer produce a significant result (p-value $>$ 0.05) and to the point where the populations would no longer being statistically differentiable (p-value $>$ 0.5). The significance in the Mann-Whitney U test is lost when the single-transiting system radius values are increased by 0.095 \rearth. This corresponds to a p-value of 0.0505 and 0.2553 for the consideration of single-transiting planets with $R_p \leq 6$ \rearth \ and the case of all single-transiting planets, respectively.
A Mann-Whitney U test p-value of 0.5037 is obtained when the smaller single-transiting planetary sample radius values are inflated by 0.214 \rearth. When the inflation of 0.214 \rearth \ is also applied in the comparison of all compact multiple planets vs all single-transiting planets, we find a p-value of 0.9222. We conclude that the offset appears robust to the choice of metric.

We hypothesize in the following sections on a plausible origin for the observed offset in radius. We focus here upon the possibility that the offset is attributable to a difference in the typical atmosphere of a planet in a single- or multiple-transiting system. In this sense, the larger silhouette is due to atmosphere alone: whether atmospheres are likelier to be present, or simply more extended, could provide a possible explanation. Dependent upon orbital configuration (and particularly whether planets reside in orbital resonance), tidal heating could provide a mechanism for both the creation of the secondary atmosphere, and additional heat flux to increase its scale height. These possibilities are explored further in Sections \ref{sec:4.1} and \ref{sec:4.2}. In an alternative hypothesized scenario, the difference in apparent radius is not due to atmosphere but rather an offset in the typical size of the bulk planetary core. If the planet formation channels for these two subsamples were sufficiently different, they might result in a different typical planet size between apparent singles and multis. Such hypotheses have been explored in, e.g., \cite{Dawson16} and \cite{Moriarty16}, and might be attributable to either a different distribution of material in the disk, or different timescale for gas dissipation.
\ben{\cite{Chance22} found a larger rate of giant impacts for single-transiting systems which could produce an offset in the radius between single- and multiple-transiting systems.} \cite{He23} quantified an impact upon inner system architecture from the presence of an outer giant planet; potentially the presence of an outer companion could also change planet size as well as planetary spacing. Single-transiting systems may have an unknown outside companion in their system which constrains the growth of the inner transiting planet. The possibility of the single-transiting systems having $N \geq 1$ \ more undiscovered planets is unknown for the planets in our sample, and a large-scale radial velocity program would be necessary to constrain any outer companions. 

\subsection{Planets in Resonances}
\label{sec:4.1}
Planets in mean-motion resonance with each other can induce tidal forces on one another that modeling has suggested can potentially cause a $\sim10\% - 50\%$ radius inflation \citep{Millholland19, Millholland20}. To investigate the impact of resonance on planetary size, we divided our compact multiple planetary sample into planets in a 3:2 or 2:1 mean-motion resonance and those that do not exhibit a 3:2 or 2:1 resonance. Table \ref{table:resonance} breaks down the orbital demographics of the planets in multi-transiting systems in our planetary sample.

\begin{table}[ht]
\caption{Categorization of Compact Multiple Planets by Resonances} 
\centering 
\begin{tabular}{c c} 
\hline\hline 
Type of Resonance & Number of Planets \\ [0.5ex] 
\hline 
3:2 & 18 \\ 
2:1 & 34 \\
Not in a 3:2 or 2:1 & 90 \\ [1ex] 
\hline 
\end{tabular}
\label{table:resonance} 
\end{table}

\begin{figure}[ht]
    \centering
    \includegraphics[width=0.5\linewidth]{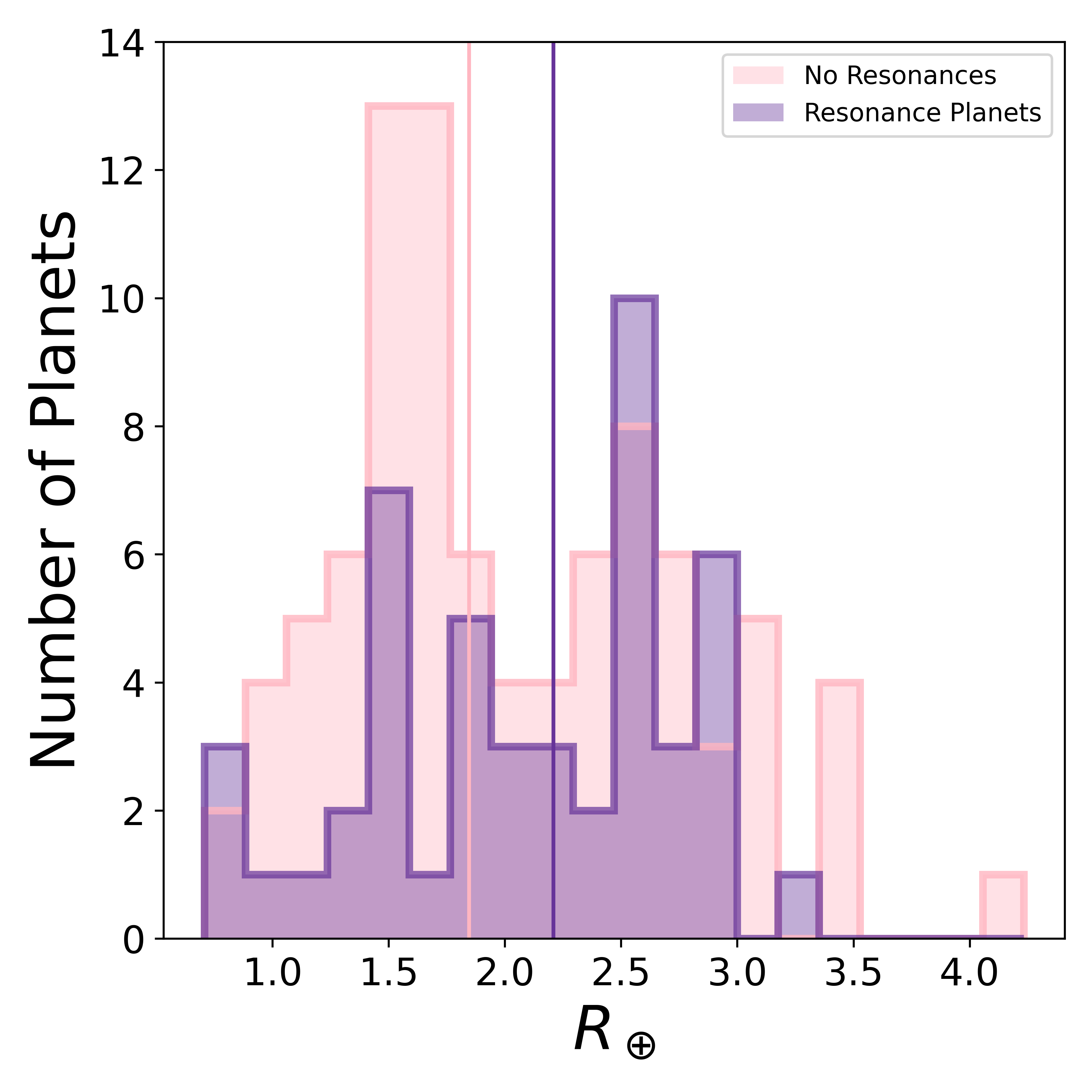}
    \caption{Overlaid histograms of compact multiple planets found in 3:2 and 2:1 orbital resonances with at least one other planet in their planetary system (purple) and all remaining other remaining compact multiple planets post quality cut (pink). The median of each sample is plotted as a vertical line of the same color. There are not enough planets which are in resonance orbits in our sample to make a robust statistical conclusion on if orbital resonant systems inhabit larger planets, and we do not find a significant difference between the population of compact multiple planets in a 3:2 or 2:1 resonance vs. those not exhibiting those desired resonances.}
    \label{fig:resonance_hist}
\end{figure}

In Figure \ref{fig:resonance_hist} we plot a histogram of the planets found in 3:2 and 2:1 resonances together vs the multi-transiting planets not found in 3:2 or 2:1 resonances. The planets labeled as not in resonance simply refers to not being in a 3:2 or 2:1 resonance. We chose to only focus on 3:2 and 2:1 since other period ratios do not show a radius inflation in models $> 2.0 \sigma$ above unity \citep{Millholland19}. While the planets which are found in resonance have a median radius value of $R_p = 2.208$ \rearth \ and the subsample of planets not in a resonance have a median radius value of $R_p = 1.845$ \rearth, we must still investigate whether this difference is statistically significant with the planet population that we have constructed.

A two-sample K-S test, Mann-Whitney U test, and Anderson-Darling k-sampling test were performed on the resonance and no-resonance planets. The results of these tests are shown in Table \ref{tab:resonance_stat_tests}. We find that there are not enough planets which are in resonance orbits in our sample to make a robust statistical conclusion, and we do not find a significant difference between the population of compact multiple planets in a 3:2 or 2:1 resonance versus those not exhibiting those desired resonances. In future work, augmenting our sample of planets with the $TESS$ sample (as long as their physical parameters are uniformly derived) may illuminate this issue. However, the $Kepler$ sample (which is designed to be a statistical sample here) is insufficiently \ben{large} in this parameter space to obtain further robust conclusion about these populations of planets.

\begin{table}[ht]
    \centering
    \caption{Statistical Tests for Planets in Resonances}
    \begin{tabular}{ccccc} 
    \hline
    \hline
         Statistical Test&  p-value & Statistic\\ 
         \hline
         Two-sample K-S Test &  0.2670 & 0.1717\\ 
         Mann-Whitney U Test& 0.3095 & 1974.0 \\ 
         Anderson-Darling K-sampling Test & 0.250 & -0.021\\ 
    \hline
    \end{tabular}
    \label{tab:resonance_stat_tests}
\end{table}

\subsection{Degassing Planetary Interiors}
\label{sec:4.2}
The dominant architectural difference between the single planetary systems and the compact multiple planetary systems is that, by definition, more orbital frequencies are relevant: this potential for resonant interaction could manifest as a higher likelihood of tidal dissipation (e.g., an increased stress on their interiors at conjunction). This extra stress and strain on the interior structure could, in turn, contribute to more efficient out-gassing of a secondary atmosphere \cite{Millholland19}. Here we investigate whether more efficient out-gassing of H$_2$O into the atmosphere may contribute to or explain the differences in the observed radii distributions between the single-transiting planets and the compact multiple-transiting systems around low mass stars. 

We assume that the bulk water content of our single-transiting planets are identical to that of the Earth, with a weight percentage \ben{(wt.\%)} of water of 0.27\% \citep{watercontentearth}.  We adopt the mass-radius relation of \citet{Zengmodel} for an Earth-like bulk composition consisting solely of iron and magnesium silicates. We note here that these models are calculated up to a planetary radius of 3 R$_\oplus$, and therefore, our single-planet sample for this analysis is limited to 97 systems out of 119 for this exercise. Here we are concerned  with the bulk distribution of the radii of the rocky planet population below this model radius and not with the tail of the distribution (or the handful of single-transiting systems containing gas giant planets); therefore this is sufficient to determine whether outgassing is a plausible mechanism to explain the differences in the radius distributions.

To determine the maximal ``atmospheric radius" of a steam atmosphere, we follow the procedure of \citet{atmospheremodelwater}, assuming 100\% of the water available in the planet is efficiently outgassed into a secondary steam atmosphere of the planet. We assume a total bulk water composition of 0.27 wt.\% \citep{watercontentearth} and calculate the extent of the steam atmosphere using the model (Equations (2) and (3)) from \citet{atmospheremodelwater}. This model also accounts for the stellar irradiation received by the planet from the host star, and therefore planets with the same planetary radius may have a different atmospheric extent than each other depending on the effective flux received from their host star.

\begin{figure}[ht]    
    \centering
    \includegraphics[width=0.45\linewidth]{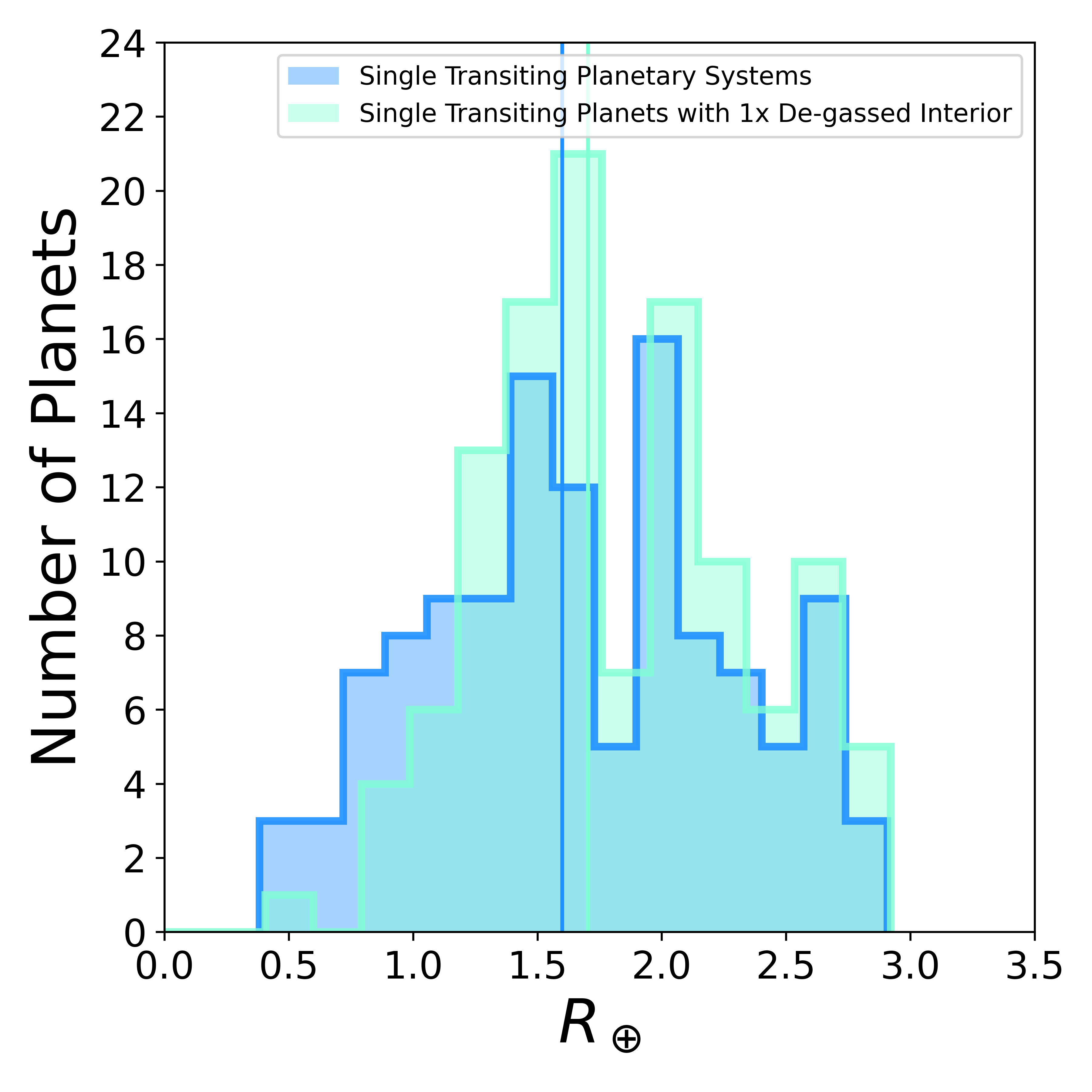}
    \includegraphics[width=0.45\linewidth]{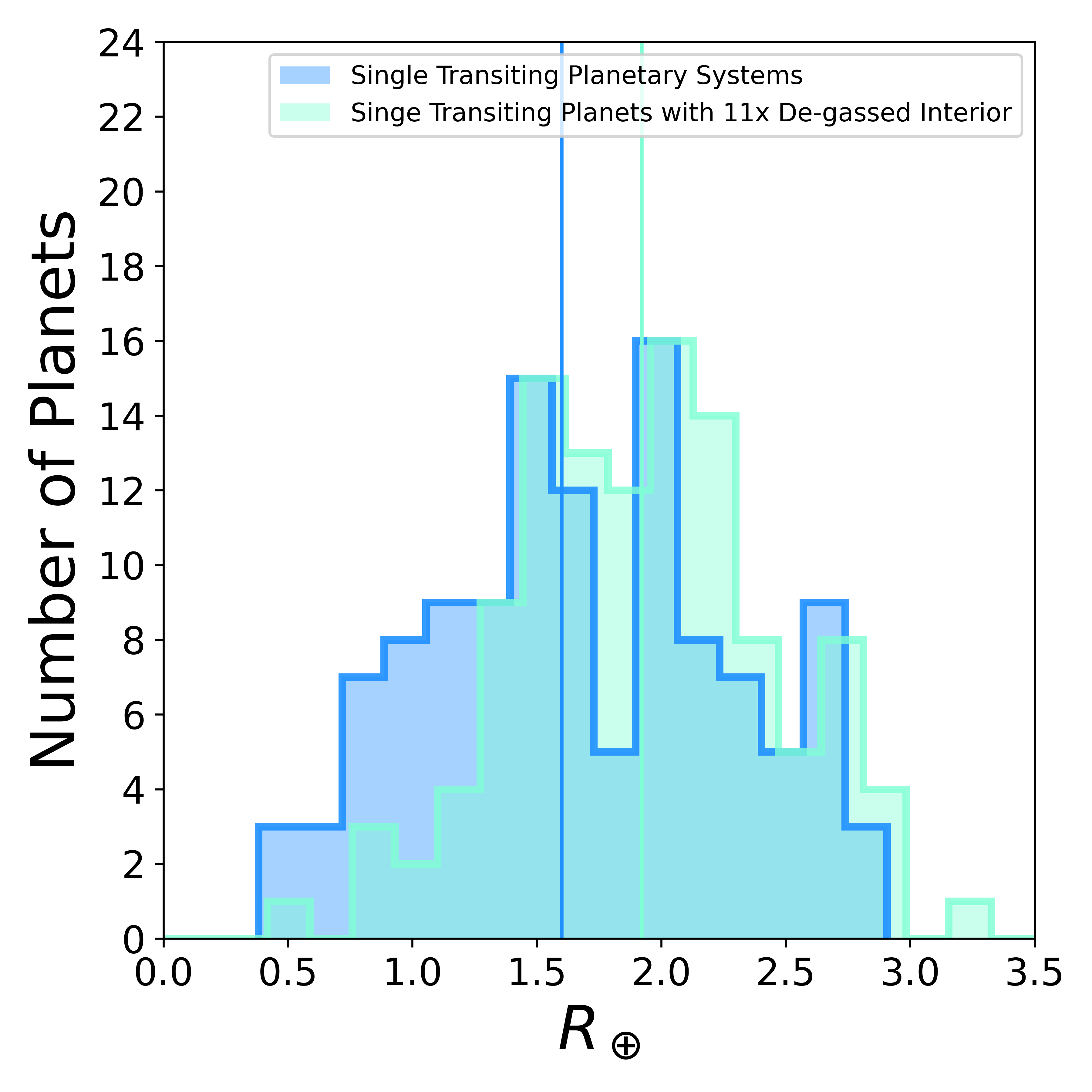}
    \caption{\ben{\textbf{Left:} Histogram of our single-transiting-planet systems (blue) and a histogram of those same planets with a steam atmosphere consisting of 0.27 wt.\% of the planet's mass (aquamarine). We estimate the planets' mass with the mass-radius relation calculated by \citet{Zengmodel} and the extent of the steam atmosphere with the atmospheric model of \cite{atmospheremodelwater}. The blue and aquamarine vertical lines represent the median of the sample. We find with 1 bulk Earth-content of water, including the surface water (the physical oceans) and the amount of water in the mantle, degassed from the interior to form a secondary atmosphere that the radii of these planets are inflated by 0.1 Earth radii, insufficient to explain our observed difference between the single and compact multiple planetary systems. \textbf{Right:} Identical to left but for a bulk water content inflated by a factor of 11 relative to the Earth. We find that to reproduce the observed differences between compact multiple planetary radii and single planetary radii through excess degassing from tidal forces that these planets would have to have a bulk water content 11$\times$ that of the Earth and for this water to be 100\% degassed to form a secondary steam atmosphere.} }
    \label{fig:water_atmosphere}
\end{figure}

In Figure \ref{fig:water_atmosphere} we show the measured radius distribution of our transiting planet systems in blue, and then with a 100\% efficiently degassed water atmosphere on top in aquamarine. We find that if we degas an Earth-composition amount of water entirely from the interior into a steam atmosphere (to produce an upper limit on how much ``extra" secondary atmosphere a compact multiple planet system might generate), we find that the median radius of our single planetary sample increases in radius by only 0.09 R$_\oplus$. \ben{To make our samples statistically indistinguishable, we require to inflate the single-transiting planets by 0.214 \rearth. Degassing the entire water content of the Earth's oceans is insufficient to account for this value, nor the much larger actual measured difference between the two samples.} If we increase the percent mass of water in these planets by a factor of 11 above the Earth value, we increase the radius distribution of our single-planet systems to a value comparable to that observed for the compact multiple-transiting planet systems. Therefore, although excess degassing of the interior from interplanetary tidal forces can partially explain the observed differences between the single planetary systems and compact multiple planetary systems, this mechanism is insufficient to fully account for these observed differences. 

Whether rocky planets can form with 11$\times$ the Earth's bulk water content by weight, even if it could be efficiently degassed, is an open question. Recent work by \citet{2023waterworlds} has found that some planets in the rocky regime, like 1.5 \rearth \ planet Kepler-138d, contain up to 11\% by mass in volatiles (dominantly water). Therefore, bulk water content in this regime is feasible; however, this would still imply different formation channels between the single-transiting-planet systems and the compact multiple systems in order to generate these vast bulk compositional differences. \ben{We suggest that the observed radius distributions of these two samples of planets across differing planetary architectures are the result of a physical difference of their bulk compositions rather than their atmospheric properties and therefore may provide clues into the formation channels of these systems.}

\subsection{Architectural Orbital Period Distribution Differences}
\ben{A natural question that arises from the consideration of the radius distribution offset in the population of single-transiting and compact multiple planetary systems is whether the orbital period distributions are also offset. \cite{Zawadzki22} evaluated whether migration traps could be the cause of the ``$Kepler$ dichotomy." \cite{Zawadzki22} proposed that clustering of planets based on size and separation of planetary semimajor axes could introduce a bias into the planetary systems of $Kepler$ which are not sensitive to planets that exist in a parameter space which avoided inward migration. To further evaluate the effect of migration traps on our sample, we plot the orbital period distributions of our planets surviving the quality cut in Section \ref{sec:3.3} in Figure \ref{fig:orb_period_hist}. In the left panel of Figure \ref{fig:orb_period_hist}, we plot the entire orbital distribution of planets surviving our quality cut. However, due to the sensitivity to shorter-period planets in our sample we zoom into the $P_{orb} < 50$ days regime in the right panel. In the right panel, the solid median lines indicate that the average orbital period of a planet in a compact multiple system is longer, on average, than a planet in a single-transiting system. In Tables \ref{tab:period_stat_tests} and \ref{tab:period_50_stat_tests} we show that the orbital period distributions of single-transiting and compact multiple planets are significantly offset with a p-value of $4.41 \times 10^{-5}$.

\begin{figure}
    \centering
    \includegraphics[width=0.45\linewidth]{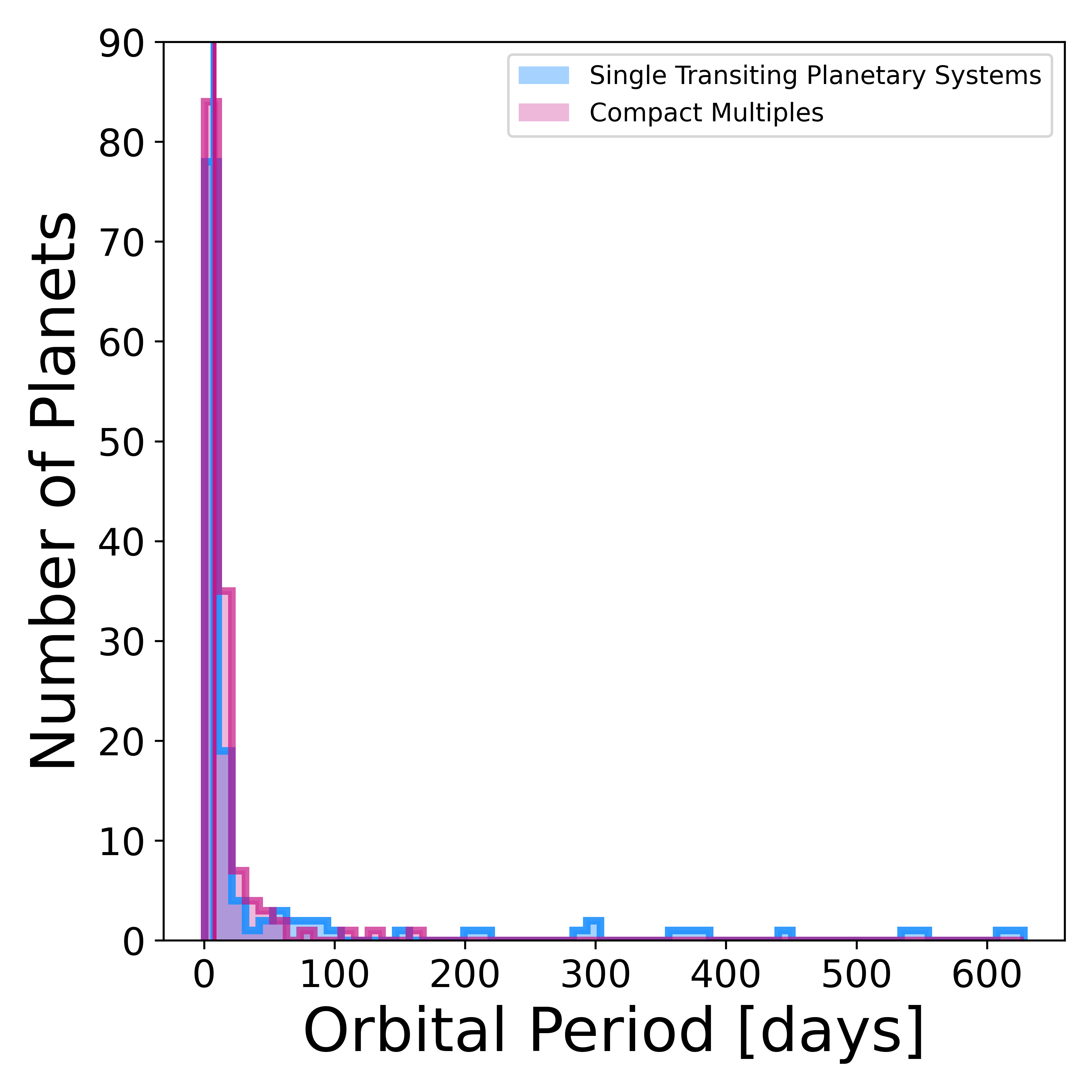}
    \includegraphics[width=0.45\linewidth]{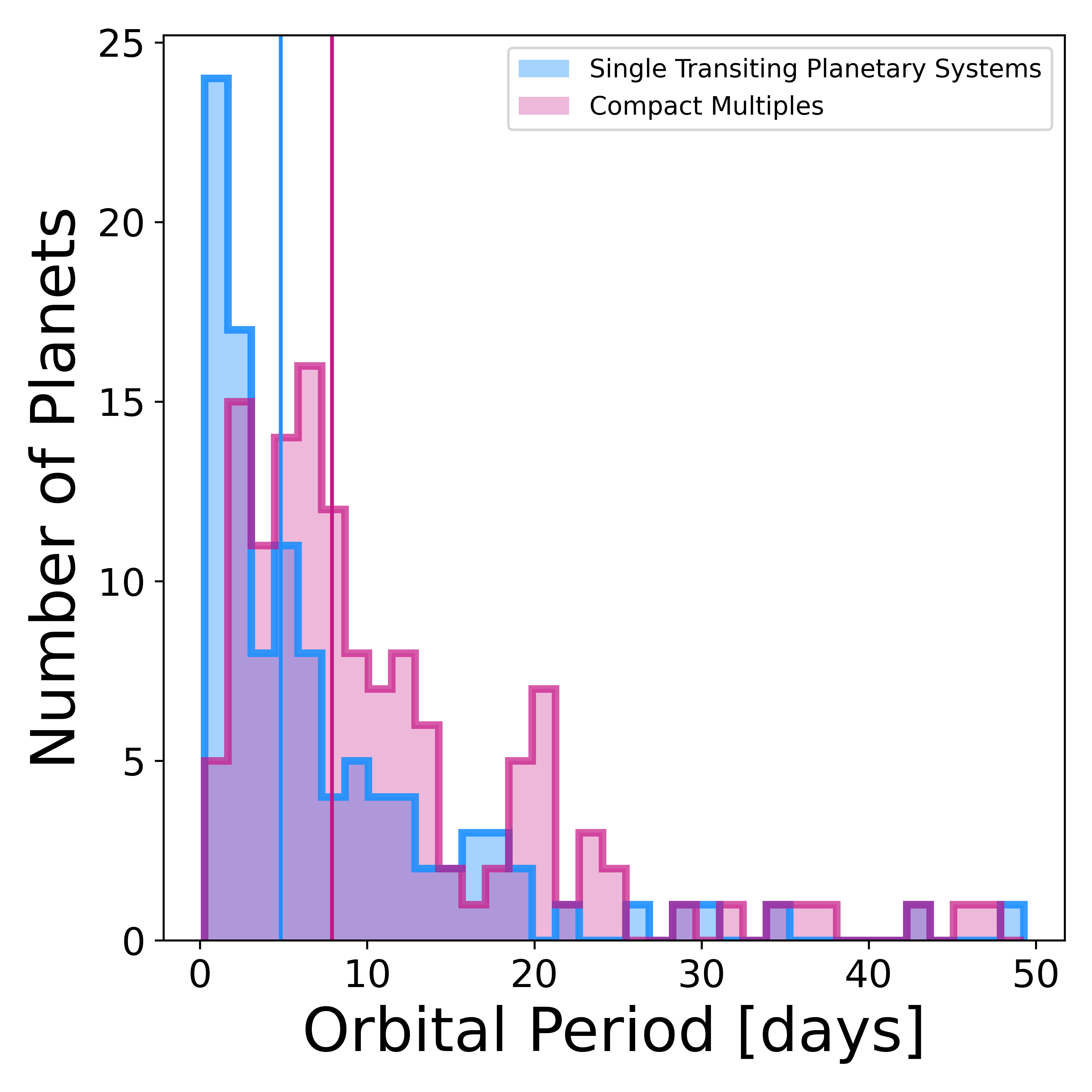}
    \caption{\ben{\textbf{Left:} Overlaid histograms of the orbital period distributions for planets in single-transiting systems (blue) and compact multiple planets (pink). We plot the median period of each sample as a solid vertical line of the respective color. The single-transiting planets have a median period of 6.29 days. The compact multiples have a median period of 8.04 days. \textbf{Right:} Overlaid histograms of the orbital period distributions for planets with orbital periods less than 50 days. This panel contains 104 single-transiting planets and 133 planets from a compact multiple system. The median of the single-transiting systems is 4.81 days, while the median of the compact multiples is 7.88 days. In both panels the compact multiples have a longer average orbital period. In Tables \ref{tab:period_stat_tests} and \ref{tab:period_50_stat_tests}, we display the results of statistically testing the difference in the orbital period distributions. With a p-value of $4.41 \times 10^{-5}$, we find that the orbital period distributions are significantly offset from each other.} }
    \label{fig:orb_period_hist}
\end{figure}

\begin{table}[ht]
    \centering
    \caption{Statistical Tests for Orbital Periods}
    \begin{tabular}{ccccc} 
    \hline
    \hline
         Statistical Test&  p-value & Statistic\\ 
         \hline
         Two-sample K-S Test &  0.0089 & 0.1980\\ 
         Mann-Whitney U Test& 0.1925 & 8074.0 \\ 
         Anderson-Darling K-sampling Test & 0.001 & 7.330\\ 
    \hline
    \end{tabular}
    \label{tab:period_stat_tests}
\end{table}

\begin{table}[ht]
    \centering
    \caption{Statistical Tests for Orbital Periods $<$ 50 days}
    \begin{tabular}{ccccc} 
    \hline
    \hline
         Statistical Test&  p-value & Statistic\\ 
         \hline
         Two-sample K-S Test &  0.0004 & 0.2664\\ 
         Mann-Whitney U Test& $4.41 \times 10^{-5}$ & 4776.0 \\ 
         Anderson-Darling K-sampling Test & 0.001 & 11.610\\ 
    \hline
    \end{tabular}
    \label{tab:period_50_stat_tests}
\end{table}

The initial shape of the single-transiting orbital period distribution in Figure \ref{fig:orb_period_hist} poses the question if our radius distribution result may be artificially created from a possible overwhelming amount of isolated hot Earths in our single-transiting sample. These isolated hot Earths represent either truly single planetary systems or dynamically isolated planets in multiplanet systems where the other planets were not sensitive to $Kepler$ detection \citep{Steffen16}. \cite{Schlaufman10} predicted $Kepler$ to find a significant number of hot super-Earths which may bias demographic studies of the full super-Earth population. Additionally, the occurrence rate of ultrashort-period planets-0.2 days to 1 day-is higher for cooler stars than hotter stars \citep{SanchisOjeda14}. To address the short period hot Earth concern, we reperform our radius analysis utilizing planets with orbital periods greater than 1.5 days. Making a period cut reduced our sample to 110 single-transiting planets and 136 compact multiple planets. In Figure \ref{fig:rad_hist_nohotearths}, we display the radius distributions of these planets with $R_p \leq 6$ \rearth \ and $P_{orb} > 1.5$ days. The median radius value of both the single-transiting and compact multiple populations increased with making a period cut to 1.884 \rearth \ and 2.013 \rearth \ respectively. The medians are offset by 0.128 \rearth. Despite the median offset being one-third of the the offset seen in Figure \ref{fig:quality_planet_hist}, we still observe a significant p-value of 0.0469 in the Mann-Whitney U test in Table \ref{tab:nohotearths_rad}. We conclude that our observed result of planets in compact multiples being larger, on average, than their single-transiting counterparts is not a result of hot Earths biasing our sample.

\begin{figure}
    \centering
    \includegraphics[width=0.5\linewidth]{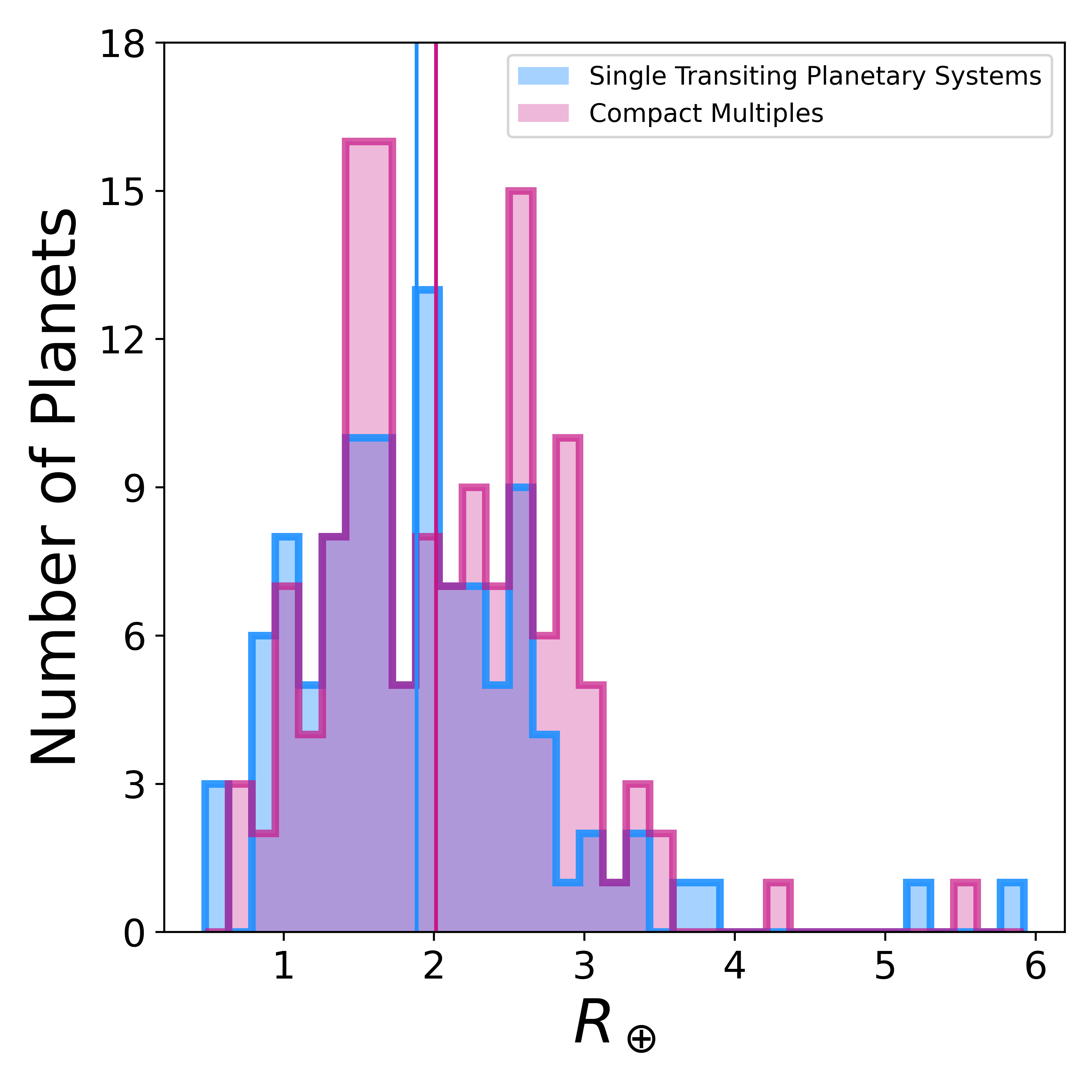}
    \caption{\ben{Overlaid histograms of the planetary radius distributions for planets surviving the quality cut in single-transiting systems (blue) and compact multiple systems (pink) with radius values less than 6 \rearth \ and orbital periods greater than 1.5 days. We plot 110 single-transiting planets and 136 planets from compact multiple systems. The solid lines representing the median of the respective populations are located at 2.013 \rearth \ and 1.884 \rearth \ for compact multiple and single-transiting planets respectively. Table \ref{tab:nohotearths_rad} contains the results from running a two-sample K-S test, Mann-Whitney U test, and Anderson-Darling k-sampling test on our reduced sample. We still obtain a significant Mann-Whitney U test p-value when making a period cutoff greater than 1.5 days in our sample.}}
    \label{fig:rad_hist_nohotearths}
\end{figure}

\begin{table}[ht]
    \centering
    \caption{Statistical Tests for Quality Planets with $R_p \leq 6$ \rearth \ \& $P_{orb} > 1.5$ days}
    \begin{tabular}{ccccc} 
    \hline
    \hline
         Statistical Test&  p-value & Statistic\\ 
         \hline
         Two-sample K-S Test &  0.1144 & 0.1499\\ 
         Mann-Whitney U Test& 0.0469 & 6377.0 \\ 
         Anderson-Darling K-sampling Test & 0.060 & 1.786\\ 
    \hline
    \end{tabular}
    \label{tab:nohotearths_rad}
\end{table}

}

\section{Conclusion}
\label{sec:conclusion}
Previous work has found that compact multiple planetary systems form more readily or have a higher probability of survival around metal poor stars \citep{Brewer18, Anderson21}. This finding is complicated by the \textit{overall} higher abundance of small planets amount metal-rich late-type stars \citep{Lu20}. Here we investigate whether there is also a difference in the radius distribution of the planets in these systems. The stellar sample used in \cite{Anderson21} was carefully chosen to be uniform and filtered out red giants from a low mass, low temperature stellar sample of late K and M dwarfs and our work is an extension of this work and based on this sample. The selection of the planet-hosting stars in the \cite{Anderson21} sample, minus the false positive systems, provided our uniform stellar sample. We determined stellar parameters in a consistent manner using \cite{Berger23} and \cite{Latham05} to acquire stellar masses and radii. Stellar radius values were used to calculate radius values for our planetary sample when combined with the measured transit depth from $Kepler$.

We performed a two-sample K-S test, Mann-Whitney U test, and Anderson-Darling k-sampling test on the compact multiple and single-transiting samples. We compared compact multiples to our bulk single-transiting planets and to the subsample of single-transiting planets with $R_p \leq$ 6 \rearth. We find that planets in compact multiples stem from a different planetary population than those in single-transiting systems in all statistical tests. \ben{\cite{RodriguezMartinez23} showed evidence that the bulk densities of planets in compact multiple systems are higher on average than in single-transiting systems. Additionally, \cite{RodriguezMartinez23} report that single-transiting planets, on average, have a higher core mass fraction (CMF) than planets in compact multiples.} Our results are consistent with those of \cite{RodriguezMartinez23}, although \cite{RodriguezMartinez23} work is derived from the RV sample of planets and primarily probes larger radii than those presented here. \ben{The decreased CMF of the planets in compact multiples is consistent with the possibility that these planets may have more volatiles which could have been outgassed from their interiors to build up a secondary atmosphere around the planets that inflate their radius values. We hint that these compositional differences are indicative of compact multiples stemming from different underlying planet formation and evolutionary channels than their single-transiting counterparts.} 

\section*{Acknowledgement}
We thank Ravit Helled, \ben{Jason Steffen, and Sarah Millholland} for useful conversations and suggestions in preparing this manuscript. \ben{We also thank our anonymous reviewer for helpful feedback in the revision of this paper.}
B.T.L. acknowledges University of Florida's University Scholars Program, which helped to support this work.
J.A.D. acknowledges the support of the Heising-Simons Foundation, whose support via the 51 Pegasi b Postdoctoral Fellowship program enabled the research leading to this project and laid the foundation for his continued professional involvement in the field.

This paper includes data collected by the Kepler mission and obtained from the MAST data archive at the Space Telescope Science Institute (STScI). Funding for the Kepler mission is provided by the NASA Science Mission Directorate. STScI is operated by the Association of Universities for Research in Astronomy, Inc., under NASA contract NAS 5–26555. 

This research has made use of the NASA Exoplanet Archive, which is operated by the California Institute of Technology, under contract with the National Aeronautics and Space Administration under the Exoplanet Exploration Program.

This work presents results from the European Space Agency (ESA) space mission Gaia. Gaia data are being processed by the Gaia Data Processing and Analysis Consortium (DPAC). Funding for the DPAC is provided by national institutions, in particular the institutions participating in the Gaia MultiLateral Agreement (MLA). The Gaia mission website is https://www.cosmos.esa.int/gaia. The Gaia archive website is https://archives.esac.esa.int/gaia. 

This work made use of the https://gaia-kepler.fun crossmatch database created by Megan Bedell.

This publication makes use of data products from the Two Micron All Sky Survey, which is a joint project of the University of Massachusetts and the Infrared Processing and Analysis Center/California Institute of Technology, funded by the National Aeronautics and Space Administration and the National Science Foundation. 

\bibliography{bibliography.bib}


\end{document}